\documentclass[preprint]{elsarticle}

\usepackage{lineno,hyperref}
\usepackage{gensymb}
\usepackage{amsmath}
\usepackage{booktabs,caption}
\usepackage[flushleft]{threeparttable}
\usepackage[version=4]{mhchem}
\usepackage[justification=centering]{caption}
\usepackage{xcolor}
\usepackage{graphicx}
\usepackage{multicol}
\usepackage{dcolumn}
\usepackage{multirow}
\usepackage{mathrsfs}
\usepackage{xspace}

\DeclareSymbolFont{matha}{OML}{txmi}{m}{it}
\DeclareMathSymbol{\varv}{\mathord}{matha}{118}
\newcommand{\mrm}[1]{\ensuremath{\mathrm{#1}}}
\newcommand{\wn}{cm$^{-1}$\xspace}

\journal{Journal of Quantitative Spectroscopy \& Radiative Transfer}

\begin{document}
	
	\begin{frontmatter}
		
		\title{High-resolution millimeter-wave spectroscopy of \ce{CH2DCl}:
			paving the way for future astronomical observations of chloromethane isotopologues}
		\tnotetext[mytitlenote]{Supplementary material available.}
		
		\author[ciamician]{Mattia Melosso\corref{mycorrespondingauthor}}
		\cortext[mycorrespondingauthor]{Corresponding author}
		\ead{mattia.melosso2(at)unibo.it}
		\author[ciamician]{Andr\`{e} Achilli}
		\author[industriale]{Filippo Tamassia}
		\author[industriale]{Elisabetta Can\`{e}}
		\author[venezia]{Andrea Pietropolli Charmet}
		\author[venezia]{Paolo Stoppa}
		\author[ciamician]{Luca Dore}
		
		\address[ciamician]{Dipartimento di Chimica ``Giacomo Ciamician'',  Universit\`a di Bologna, Via F.~Selmi 2, 40126 Bologna (Italy)}
		\address[industriale]{Dipartimento di Chimica Industriale ``Toso Montanari'', Universit\`a di Bologna, Viale del Risorgimento 4, 40136 Bologna (Italy)}
		\address[venezia]{Dipartimento di Scienze Molecolari e Nanosistemi, Universit\`a Ca' Foscari Venezia, Via Torino 155, 30172 Mestre (Italy)}
		
		
		\begin{abstract}
			Chloromethane is the only organochloride detected in space to date.
			Its recent observation towards the low-mass protostar IRAS 16293-2422 with ALMA offers
			a prompt for new laboratory studies of \ce{CH3Cl} and its isotopologues.
			Here, we report the investigation of the rotational spectrum of monodeuterated chloromethane
			\ce{CH2DCl} in the frequency region between 90 and 300\,GHz.
			The measurements have been carried out with a frequency-modulation millimeter-wave spectrometer,
			arranged to perform saturation spectroscopy.
			From the analysis of hyperfine-resolved spectra of the two chlorine isotopologues \ce{CH2D^{35}Cl} and
			\ce{CH2D^{37}Cl}, consistent sets of accurate spectroscopic parameters have been obtained.
			This work provides reliable spectral predictions which can be used to guide radio-astronomical searches
			of \ce{CH2DCl} in the interstellar medium and represents a solid base for future analyses of
			high-resolution infrared spectra of monodeuterated chloromethane.
			
		\end{abstract}
		
		
		
	\end{frontmatter}

	
	\section{Introduction}\label{sec:intro}
	
	\noindent
	In the family of halocarbons, chloromethane (\ce{CH3Cl}) represents the smallest member of the chlorine-bearing species.
	This organochloride is massively-employed in industry as a methylating and chlorinating agent, while its use as refrigerant (R-40) ceased because of its ozone depletion potential and global warming power \cite{emani2017development}.
	Indeed, large amounts of \ce{CH3Cl} are known to be present in the Earth's atmosphere: its atmospheric abundance has been demonstrated to be due to both anthropogenic (accidental releases or burning processes) and natural causes (produced by plants, bacteria, phytoplankton) \cite{keppler2005new}.
	
	\noindent
	Atmospheric \ce{CH3Cl}, previously revealed by \emph{in situ} measurements \cite{lovelock1975natural,grimsrud1975survey}, has been spectroscopically identified for the first time in the solar absorption spectrum of our troposphere through its strong infrared (IR) features around 2967\,\wn \cite{park1986spectroscopic}, corresponding to the $Q$ branch of the $\nu_1$ vibrational mode.
	Because of its tendency to accumulate in the upper levels of the atmosphere, chloromethane has been suggested as an optimal biosignature gas to be searched for in potentially habitable exoplanets \cite{seager2016toward,schwieterman2018exoplanet}.
	
	\noindent
	The presence of chloromethane in environments other than our Solar system has been recently identified at a pre-planetary stage, thanks to the observation of \ce{CH3Cl} emission towards the low-mass protostar IRAS 16293-2422 with the Atacama Large Millimeter/submillimeter Array (ALMA) \cite{fayolle2017protostellar}.
	In particular, this detection relied on the observation at submillimeter-wavelengths of some $J = 13 \leftarrow 12$ rotational transitions of the two stable chlorine isotopologues, \ce{CH3{}^{35}Cl} and \ce{CH3{}^{37}Cl}.
	
	\noindent
	The protostar IRAS 16293-2422 is also a rich source of interstellar deuterated species \cite{van1995molecular}.
	During the last 25 years, a plethora of deuterium-bearing molecules have been identified in this source, including multiply-deuterated forms of water (\ce{HDO} and \ce{D2O}) \cite{coutens2012study}, ammonia (up to \ce{ND3}) \cite{roueff2005interstellar}, methanol (up to \ce{CD3OH}) \cite{parise2004first}, and methyl cyanide (\ce{CH2DCN} and \ce{CHD2CN}) \cite{calcutt2018alma}.
	High abundances of many other deuterated species are predicted by astrochemical models, which unfortunately have not yet taken into account deuterium enrichment in \ce{CH3Cl}.
	However, given its chemical similarity with \ce{CH3CN} and \ce{CH3OH}, \ce{CH3Cl} can likely exhibit a strong deuterium fractionation (or D/H ratio) in IRAS 16293-2422, thus producing significant amounts of monodeuterated chloromethane \ce{CH2DCl}.
	
	\noindent
	A vast spectroscopic literature is available for the parent species \ce{CH3Cl} about its vibrational
	and rotational spectra (see Refs.~\cite{nikitin2016improved,stvriteska2009precise} and references therein).
	Extended line lists \cite{owens2018exomol} and molecular database \cite{gordon2017hitran2016} are available for
	both the \ce{CH3{}^{35}Cl} and \ce{CH3{}^{37}Cl} species.
	Also the rare \ce{^{13}C} isotopologues have been spectroscopically characterized  widely, from the millimeter-wave to the infrared domains (see, e.g., Refs.~\cite{litz2003infrared,kania2008rotational} and references therein).
	On the other hand, deuterated forms of \ce{CH3Cl} have been poorly characterized from a spectroscopic point-of-view.
	
	\noindent
	Albeit the fundamental $J_{K_a,\,K_c} = 1_{\,0,\,1} \leftarrow 0_{\,0,\,0}$ rotational transition of \ce{CH2DCl} was recorded in 1950 by Stark modulation microwave spectroscopy \cite{matlack1950microwave} and in the early 1970s with a molecular-beam maser spectrometer \cite{kukolich1971high,kukolich1972variation}, the rotational spectrum of \ce{CH2DCl} remains substantially unstudied so far.
	Since most of the molecules observed in space are detected through rotational lines emission \cite{mcguire2018census}, the lack of such spectral data poses substantial limitations to the identification of monodeuterated chloromethane in the interstellar medium (ISM).
	As far as the infrared spectrum of \ce{CH2DCl} is concerned, the analysis of six fundamental and some of their hot-bands has been recently reported for the \ce{^{35}Cl} isotopologue \cite{baldacci2005high,baldacci2008high,baldacci2010high}, while no data are available for the \ce{^{37}Cl} one.
	
	\noindent
	Here, we report the investigation of the rotational spectra of \ce{CH2D^{35}Cl} and \ce{CH2D^{37}Cl}, observed at millimeter-wavelengths with a frequency-modulation (FM) absorption spectrometer.
	This work is aimed at (i) providing reliable spectral predictions that will assist radio-astronomical searches of \ce{CH2DCl} and (ii) obtaining a set of accurate spectroscopic parameters for both the chlorine isotopologues, which constitutes a solid base to further investigate the infrared spectrum of monodeuterated chloromethane.
	
	
	\section{Experimental details}\label{sec:exp}
	
	\noindent
	A pure sample of \ce{CH2DCl} was synthesized in our laboratory by reacting monodeuterated methanol (\ce{CH2DOH}; CND Isotopes, 99.2\% D-enriched) with sodium chloride (\ce{NaCl}) in acid aqueous solution, following the procedure described in Ref.~\cite{baldacci2005high}.
	
	\noindent
	Rotational spectra of \ce{CH2DCl} were recorded in the frequency ranges 90--125\,GHz and 240--300\,GHz
	using a FM millimeter-/submillimeter-wave spectrometer, used in past for the study of other deuterated
	species \cite{melosso2019pure,melosso2019rotational,degli2019determination}.
	
	\noindent
	Two Gunn diodes from J. E. Carlstrom Co. and Radiometer Physics GmbH, emitting between 80--115\,GHz and 116--125\,GHz, respectively, are used as primary radiation source of the instrument. Spectral coverage at higher frequencies is obtained by coupling the Gunn diodes to passive frequency multipliers, namely doublers and triplers in cascade.
	A 75\,MHz sine-wave modulated wave is used as reference signal in a Phase-Lock Loop
	through which the Gunn's radiation is locked to a harmonic of a digital synthesizer (HP8672A, 2--18\,GHz);
	in this way, the frequency modulation $f$ is transferred to the output radiation.
	The frequency accuracy of the radiation is guaranteed by locking the radio-frequency synthesizers to a rubidium atomic clock.
	
	\noindent
	The millimeter-wave is fed to a 3.25\,m long glass absorption cell, closed at the two ends by high-density polyethylene windows and connected to a pumping system. Doppler-limited spectra were recorded by filling the cell with \ce{CH2DCl} vapors at the stagnation pressure of 6\,Pa, in order to minimise pressure-broadening effects; higher pressures (up to 25\,Pa) were
	used to record the very weak $b$-type transitions (see \S\ref{sec:analysis}).
	
	\noindent
	Two Schottky barrier diodes were used as detector, from Millitech Co. up to 125\,GHz and the WR3.4ZBD from Virginia Diodes between 240 and 300\,GHz. The detected signal is pre-amplified, filtered and demodulated at $2f$ by an analog Lock-in, digitally-converted, and finally sent to a computer.
	
	\noindent
	Additional measurements in the 240--300\,GHz frequency range were performed exploiting the Lamb-dip technique \cite{lamb1964theory}.
	In this case, the optics of the spectrometer were appropriately set up in a double-pass configuration,
	as described in Ref.~\cite{melosso2020nh2d}. Also, a low-pressure ($\sim$1\,Pa) of \ce{CH2DCl}, and
	$f$ and modulation-depth values as low as 1\,kHz and 15\,kHz, respectively, were used.
	With such experimental conditions, chlorine hyperfine splittings could be well-resolved even at high frequencies.
	
	
	\section{Spectral analysis}\label{sec:analysis}
	
	\noindent
	From a spectroscopic point of view, \ce{CH2DCl} is a nearly-prolate asymmetric-top rotor belonging to
	the $C_\mrm{s}$ point group. Therefore, its nine vibrational modes are either of $A'$ or $A''$ symmetry,
	as illustrated for \ce{CH2D^{35}Cl} in Table~\ref{tab:modes}.
	
	\begin{table}[htb!]
		\centering
		\caption{Vibrational modes and their energy for \ce{CH2D^{35}Cl}.}
		\label{tab:modes}
		\scalebox{0.99}{
			\begin{threeparttable}
				\begin{tabular}{cclrc}
					\hline\hline \\[-1ex]
					Symmetry & Mode & Description & Wavenumber\tnote{a} & Reference  \\[0.5ex]
					\hline \\[-1.5ex]
					A'    &  $\nu_1$     & \ce{CH2} sym. stretching   &  2989.9(3)      & \cite{baldacci2005high}  \\[0.5ex]
					&  $\nu_2$     & \ce{C-D} stretching        &  2223.7(3)      & \cite{baldacci2005high}  \\[0.5ex]
					&  $\nu_3$     & \ce{CH2} scissoring        &  1433.839(3)    & \cite{baldacci2008high}  \\[0.5ex]
					&  $\nu_4$     & \ce{CH2} wagging           &  1268.3335(1)   & \cite{baldacci2010high}  \\[0.5ex]
					&  $\nu_5$     & \ce{C-D} in plane bending  &  827.02343(8)   & \cite{baldacci2005high}  \\[0.5ex]
					&  $\nu_6$     & \ce{C-Cl} stretching       &  714.11267(9)   & \cite{baldacci2005high}  \\[0.5ex]
					A''   &  $\nu_7$     & \ce{CH2} asym. stretching  &  3035.3(3)      & \cite{baldacci2005high}  \\[0.5ex]
					&  $\nu_8$     & \ce{C-D/CH2} out of plane bending &  1267.6775(1)   & \cite{baldacci2010high}  \\[0.5ex]
					&  $\nu_9$     & \ce{C-D/CH2} out of plane bending &  986.69013(6)   & \cite{baldacci2008high}  \\[0.5ex]
					\hline\hline
				\end{tabular}
				\begin{tablenotes}
					\item[a] Units are \wn. Numbers in parenthesis represent quoted uncertainties.
				\end{tablenotes}
			\end{threeparttable}
		}
	\end{table}
	
	\noindent
	Differently from the parent species \ce{CH3Cl} ($C_\mrm{3v}$ group), in which the carbon-halogen bond lies
	on the symmetry axis, the $a$-axis of the principal inertia system of \ce{CH2DCl} is slightly rotated
	with the respect to the \ce{C-Cl} bond because of the different center of mass.
	Replacing one hydrogen atom with deuterium causes the permanent dipole moment ($\mu=1.870$\,D for \ce{CH3Cl}
	\cite{wlodarczak1985dipole}) to be distributed along two components; based on geometric considerations
	\cite{black2001general}, and assuming that the total dipole moment does not change among the isotopologues,
	one can estimate $\mu_a=1.868$\,D and $\mu_b=0.076$\,D for \ce{CH2DCl}.
	
	\begin{figure}[htb!]
		\centering
		\includegraphics[width=0.99\textwidth]{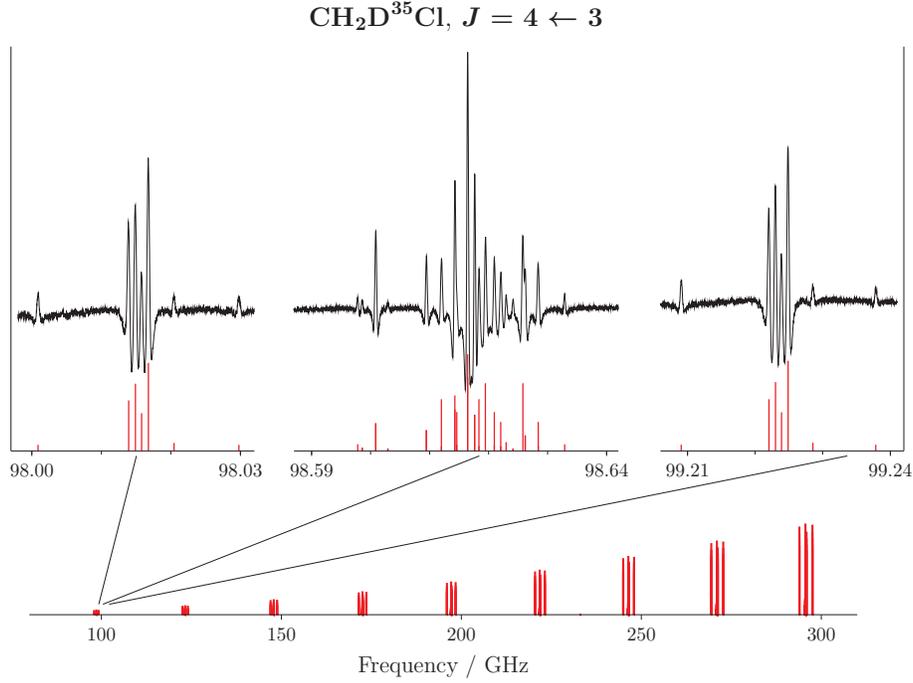}
		\caption{The hyperfine structure of the $J$ = 4 $\leftarrow$ 3 transition of \ce{CH2D^{35}Cl}. The black trace is
			the experimental spectrum, red sticks symbolise hyperfine components as predicted from the final fit. The $K_a=1$
			doublet is shown in the side spectra, whereas the central spectrum contains the $K_a=$ 0, 2 and 3 components.}
		\label{fig:over}
	\end{figure}
	
	\noindent
	The rotational energy of \ce{CH2DCl} can be modeled by using the standard semi-rigid Hamiltonian for an
	asymmetric rotor with a non-vanishing nuclear spin:
	
	\begin{equation}\label{eq:ham}
	\mathscr{H} = \mathscr{H}_\mrm{rot} + \mathscr{H}_\mrm{cd} + \mathscr{H}_\mrm{hfs} \,,
	\end{equation}
	
	\noindent
	where $\mathscr{H}_\mrm{rot}$ contains the rotational constants in the $A$-Watson reduced form \cite{watson1967determination}:
	
	\begin{equation}\label{eq:rot}
	\mathscr{H}_\mrm{rot} = \tfrac{1}{2}\left( B+C \right)\hat{P}^2 + \left[A-\tfrac{1}{2}\left( B+C \right)\right]\hat{P}_a^2 + \tfrac{1}{2}\left( B-C \right)\left(\hat{P}_b^2-\hat{P}_c^2\right) ,
	\end{equation}
	
	\noindent
	the $\mathscr{H}_\mrm{cd}$ part accounts for centrifugal distortion terms with increasing power of the angular momentum
	
	\begin{multline}\label{eq:dist}
	\mathscr{H}_\mrm{cd} = - \Delta_J\hat{P}^4 - \Delta_{JK}\hat{P}^2\hat{P}_a^2 - \Delta_K\hat{P}_a^4 \\
	- \delta_J\hat{P}^2\left(\hat{P}_b^2-\hat{P}_c^2\right)
	- \delta_K \left[\hat{P}^2\left(\hat{P}_b^2-\hat{P}_c^2\right)+\left(\hat{P}_b^2-\hat{P}_c^2\right)\hat{P}^2\right] + \ldots \,,
	\end{multline}
	
	\noindent
	while $\mathscr{H}_\mrm{hfs}$ contains the nuclear quadrupole coupling (NQC) $\chi_{ii}$ and spin-rotation (SR) $C_{ii}$
	constants of the chlorine nuclear spins ($I_\mrm{Cl}=3/2$).
	Deuterium is a quadrupolar nucleus too ($I_\mrm{D}=1$), but its contribution to the rotational energy levels is negligible
	in the millimeter region and, thus, is not considered in this work.
	
	\noindent
	Both chlorine isotopologues of \ce{CH2DCl} show a strong $a$-type spectrum, with groups of $(J+1) \leftarrow J$ transitions
	spaced by nearly $(B+C) \simeq 24$\,GHz. The structure of each transition is typical of an asymmetric rotor very close to the prolate limit: most of the $K_a$ components are grouped in the proximity of the $K_a=0$, whereas the high and low components of the $K_a=1$ doublet are found \emph{ca.} $\tfrac{1}{2}(B-C)(J+1)$ above or below, respectively.
	Moreover, because the chlorine quadrupolar interactions split each rotational level into four sub-levels with $F = J+3/2; \,J+1/2; \,J-1/2; \,J-3/2$, a hyperfine structure (HFS) is produced in the spectrum.
	Considering the selection rules $\Delta F=0;\pm1$, up to nine hyperfine components are allowed for each $J'_{K_a',\,K_c'} \leftarrow J_{K_a,\,K_c}$ transition, the strongest components being those with $\Delta F=\Delta J$.
	As an example, the complexity of the $J$ = 4 $\leftarrow$ 3 transition, including its HFS, is highlighted in Figure~\ref{fig:over}.
	
	\section{Results and Discussion}\label{cha:disc}
	
	\noindent
	Spectral predictions for \ce{CH2D^{35}Cl} were initially performed by using the ground state
	spectroscopic constants reported in Ref.~\cite{baldacci2005high} together with the NQC constants
	from Ref.~\cite{kukolich1971high}. As far as \ce{CH2D^{37}Cl} is concerned, rotational and centrifugal distortion
	constants were evaluated from the chloromethane equilibrium geometry of	Ref.~\cite{black2001general}
	and the NQC constants taken from Ref.~\cite{kukolich1972variation}.
	
	\begin{figure}[htb!]
		\centering
		\includegraphics[width=0.60\textwidth]{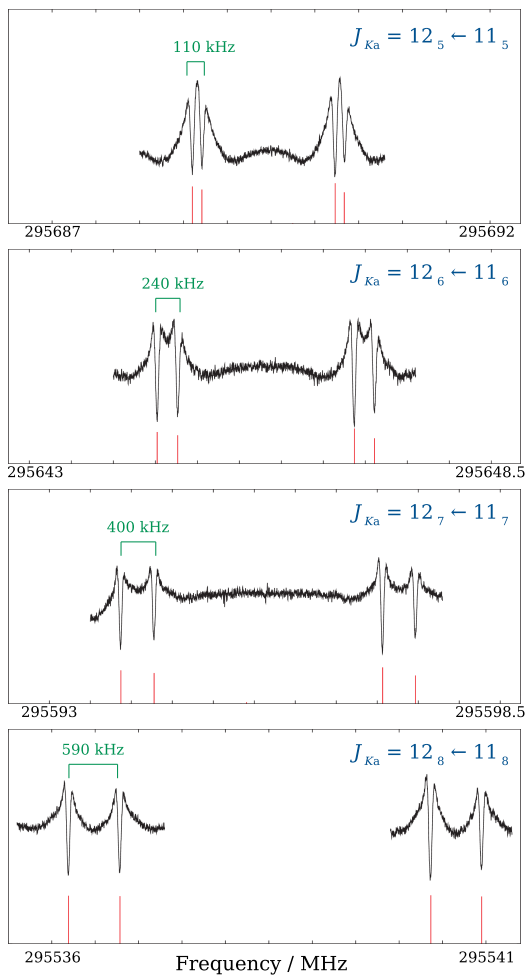}
		\caption{Splitting evolution in \ce{CH2D^{35}Cl} Lamb-dip spectra. The black trace is
			the experimental spectrum, red sticks symbolise hyperfine components as predicted from the final fit.
			The HF components in each panel are, from lower to higher frequency: $F' \leftarrow F= 12.5 \leftarrow 11.5, 11.5 \leftarrow 10.5, 13.5 \leftarrow 12.5, 10.5 \leftarrow 9.5$.
			The magnitude of some hyperfine splittings is indicated in green.
			The quantum numbers $K_c$ have been omitted in the labelling of each transition, because asymmetry splittings
			are not resolved.}
		\label{fig:split}
	\end{figure}
	
	\noindent
	First, we recorded five complete $a$-type $(J+1) \leftarrow J$ transitions for both isotopologues,
	with $4 \leq J \leq 11$. Once the sets of spectroscopic parameters had been refined and the uncertainty
	of the $A$ constants reduced, we have searched for the extremely weak $b$-type transitions.
	They were expected to be about 600 times weaker than the $a$-type ones, but thanks to the high-sensitivity
	of the spectrometer we were able to detect them with a good signal-to-noise ratio (SNR) in the spectrum.
	
	\noindent
	Also, most transitions in the frequency range 240--300\,GHz have been recorded with the Lamb-dip technique,
	thus allowing the resolution of small hyperfine splittings even at high $J$ values.
	Some Lamb-dip spectra are shown in Figure~\ref{fig:split}, in which the splitting evolution for different
	$K_a$ transitions can be noticed.
	
	\noindent
	In addition to the measurements of the ground state spectra of \ce{CH2DCl}, three R branch $a$-type
	rotational transitions have been observed for \ce{CH2D^{35}Cl} in the $\varv_5=1$ and $\varv_6=1$
	vibrational excited states, previously investigated in the infrared region \cite{baldacci2005high}.
	
	\noindent
	The data analysis has been carried out with the SPFIT subprogram of the CALPGM suite \cite{pickett1991fitting}.
	For the \ce{CH2D^{35}Cl} isotopologue, the newly measured ground and excited states transitions have been
	fitted in a weighted least-square procedure together with ro-vibrational data from the literature \cite{baldacci2005high}.
	Each datum has a weight proportional to the inverse square of its uncertainty.
	Infrared data were used with half the uncertainties quoted in the original paper, as suggested by their fit standard deviation \cite{baldacci2005high}. The experimental error of our transition frequencies, instead, was estimated to be between 15 and 50\,kHz, mostly depending on the SNR of the	observed absorption line, and around 3\,kHz for Lamb-dip measurements.
	The same criteria were applied to evaluate the experimental error associated to each transition of the \ce{CH2D^{37}Cl} isotopologue, whose analysis is based on our measurements only.
	
	\noindent
	Thanks to the observation of both $a$- and $b$-type transitions, chlorine-resolved hyperfine components, and Lamb-dip measurements,
	an accurate set of spectroscopic parameters has been attained for each \ce{CH2DCl} isotopologue.
	They include the rotational constants, the complete set of quartic and some sextic centrifugal distortion terms, the diagonal $\chi_{ii}$(Cl) constants of the NQC tensor, and the $C_{ii}$(Cl) SR constants.
	The final parameters are collected in Tables~\ref{tab:ground} and~\ref{tab:ex}, where they are compared with those of Ref.~\cite{baldacci2005high}.
	
	\begin{table*}[htb!]
		\caption{Spectroscopic parameters determined for \ce{CH2DCl} in the ground vibrational state$^{\,[a]}$.}
		\label{tab:ground}
		\centering
		\scalebox{0.9}{
			\begin{tabular}{p{1.6cm}rD{.}{.}{10}D{.}{.}{10}D{.}{.}{10}}
				\hline\hline \\[-1ex]
				Parameter & Unit & \multicolumn{1}{c}{This work} & \multicolumn{1}{c}{Previous IR$^{\,[b]}$} & \multicolumn{1}{c}{This work} \\[0.5ex]
				&      & \multicolumn{2}{c}{\ce{CH2D^{35}Cl}} & \multicolumn{1}{c}{\ce{CH2D^{37}Cl}} \\[0.5ex]
				\hline \\[-1ex]
				$            A $  &  MHz  &  119836.774(23) &   119890.(42) &   119834.890(23) \\[0.5ex]
				$            B $  &  MHz  & 12479.34832(74) & 12479.392(39) &  12278.10755(72) \\[0.5ex]
				$            C $  &  MHz  & 12177.17067(76) & 12177.117(39) &  11985.46816(51) \\[0.5ex]
				$     \Delta_J $  &  MHz  &  0.01503113(40) & 0.0150555(60) &   0.01458138(72) \\[0.5ex]
				$  \Delta_{JK} $  &  MHz  &   0.1564373(24) &  0.155233(90) &    0.1520332(85) \\[0.5ex]
				$     \Delta_K $  &  MHz  &     1.64916(94) &      2.04(33) &       1.6526(18) \\[0.5ex]
				$     \delta_J $  &  kHz  &     0.36830(42) &    0.3699(45) &      0.34809(47) \\[0.5ex]
				$     \delta_K $  &  MHz  &     0.02440(28) &    0.0330(90) &      0.02227(30) \\[0.5ex]
				$    \Phi_{J}  $  &  mHz  &       -1.46(25) &               &                  \\[0.5ex]
				$    \Phi_{JK} $  &   Hz  &     -0.0461(64) &               &       0.316(33)  \\[0.5ex]
				$    \Phi_{KJ} $  &   Hz  &      3.8555(18) &               &       3.635(28)  \\[0.5ex]
				$\chi_{aa} $(Cl)  &  MHz  &     -74.471(12) &               &    -58.703(14)   \\[0.5ex]
				$\chi_{bb} $(Cl)  &  MHz  &      37.116(24) &               &     29.294(22)   \\[0.5ex]
				$C_{aa}$(Cl)      &  kHz  &        3.9(10)  &               &        4.4(17)   \\[0.5ex]
				$C_{bb}$(Cl)      &  kHz  &        3.46(76) &               &                  \\[0.5ex]
				$C_{cc}$(Cl)      &  kHz  &        1.25(76) &               &        3.37(54)  \\[0.5ex]
				\hline \\[-1ex]
				IR data                  &              & \multicolumn{1}{c}{2708}                 & \multicolumn{1}{c}{2708}                 &                            \\[0.5ex]
				\multicolumn{2}{l}{IR $(J,\, K_{a})_\mrm{max}$} & \multicolumn{1}{c}{57, 14}               & \multicolumn{1}{c}{57, 14}               &                            \\[0.5ex]
				IR \emph{rms}            & \wn          & \multicolumn{1}{c}{4.4$\times$10$^{-4}$} & \multicolumn{1}{c}{4.9$\times$10$^{-4}$} &                            \\[0.5ex]
				MW data                  &              & \multicolumn{1}{c}{272}                  &                                          & \multicolumn{1}{c}{210}    \\[0.5ex]
				\multicolumn{2}{l}{MW $(J,\, K_{a})_\mrm{max}$} & \multicolumn{1}{c}{30, 11}               &                                          & \multicolumn{1}{c}{35, 11} \\[0.5ex]
				MW \emph{rms} & \multicolumn{1}{l}{kHz} & \multicolumn{1}{c}{17.7}                 &                                          & \multicolumn{1}{c}{15.0}   \\[0.5ex]
				$\sigma$                 &              & \multicolumn{1}{c}{0.83}                 &                                          & \multicolumn{1}{c}{0.77}   \\[0.5ex]
				\hline\hline \\[-1ex]
			\end{tabular}
		}
		\smallskip
		\\ \textbf{Notes:} \textbf{[a]} Numbers in parenthesis are one standard deviation and apply to the last significant digits.
		\textbf{[b]} Ref.~\cite{baldacci2005high}.
	\end{table*}
	
	
	\begin{table*}[htb!]
		\caption{Spectroscopic parameters determined for \ce{CH2D^{35}Cl} in the singly-excited vibrational states $\varv_5=1$ and $\varv_6=1$.$^{\,[a]}$}
		\label{tab:ex}
		\centering
		\scalebox{0.8}{
			\begin{tabular}{p{1.6cm}rD{.}{.}{10}D{.}{.}{10}D{.}{.}{10}D{.}{.}{10}}
				\hline\hline \\[-1ex]
				Parameter & Unit & \multicolumn{2}{c}{This work} & \multicolumn{2}{c}{Previous IR$^{\,[b]}$} \\[0.5ex]
				& & \multicolumn{1}{c}{$\varv_6=1$} & \multicolumn{1}{c}{$\varv_5=1$} & \multicolumn{1}{c}{$\varv_6=1$} & \multicolumn{1}{c}{$\varv_5=1$} \\[0.5ex]
				\hline \\[-1ex]
				$           E $ &  \wn  & 714.112221(24)  &  827.023678(23)   & 714.11267(9) & 827.02343(8) \\[0.5ex]
				$           A $ &  MHz  &  119734.935(32) &   119936.977(37)  & 119787.68(6) & 119990.67(9) \\[0.5ex]
				$           B $ &  MHz  & 12390.1856(21)  &  12426.52318(83)  &  12390.16(1) &  12426.56(1) \\[0.5ex]
				$           C $ &  MHz  & 12086.4828(32)  &  12127.4790(26)   &  12086.64(2) &  12127.24(3) \\[0.5ex]
				$    \Delta_J $ &  MHz  &  0.01529052(94) &   0.01480945(63)  &  0.015295(2) &  0.014827(3) \\[0.5ex]
				$ \Delta_{JK} $ &  MHz  &   0.1608453(90) &    0.151884(10)   &   0.15841(9) &   0.15247(9) \\[0.5ex]
				$    \Delta_K $ &  MHz  &     1.66109(95) &      1.62429(96)  &    2.0506(4) &    2.0159(5) \\[0.5ex]
				$    \delta_J $ &  kHz  &     0.43118(50) &      0.26756(64)  &     0.428(2) &     0.270(4) \\[0.5ex]
				$    \delta_K $ &  MHz  &     0.02948(90) & 0.02440^{\,[c]} &     0.032(2) &     0.019(3) \\[0.5ex]
				$\chi_{aa}$(Cl) &  MHz  &  -74.632(61)    &    -74.629(80)    &              &              \\[0.5ex]
				$\chi_{bb}$(Cl) &  MHz  &    39.9(24)     & 37.116^{\,[c]}  &              &              \\[0.5ex]
				\hline \\[-1ex]
				$      G_{c}  $ &  MHz  &    \multicolumn{2}{c}{4821.13(86)}  &  \multicolumn{2}{c}{4904.(9)}  \\[0.5ex]
				$ G_c^{JK}  $ &  kHz  &    \multicolumn{2}{c}{-0.2087(32)}  &  \multicolumn{2}{c}{}          \\[0.5ex]
				$      F_{ab} $ &  MHz  &    \multicolumn{2}{c}{-2.3034(98)}  &  \multicolumn{2}{c}{-1.16(1)}  \\[0.5ex]
				\hline \\[-1ex]
				IR data         &       & \multicolumn{2}{c}{2708}  & \multicolumn{2}{c}{2708} \\[0.5ex]
				\multicolumn{2}{l}{IR $(J,\, K_{a})_\mrm{max}$} & \multicolumn{2}{c}{57, 14} & \multicolumn{2}{c}{57, 14} \\[0.5ex]
				IR \emph{rms}            & \wn          & \multicolumn{2}{c}{4.4$\times$10$^{-4}$} & \multicolumn{2}{c}{4.9$\times$10$^{-4}$} \\[0.5ex]
				MW data         &       & \multicolumn{2}{c}{202}  & \multicolumn{2}{c}{} \\[0.5ex]
				\multicolumn{2}{l}{MW $(J,\, K_{a})_\mrm{max}$} & \multicolumn{2}{c}{12, 11} & \multicolumn{2}{c}{} \\[0.5ex]
				MW \emph{rms}   & kHz   & \multicolumn{2}{c}{30.4}  & \multicolumn{2}{c}{} \\[0.5ex]
				$\sigma$        &       & \multicolumn{2}{c}{0.83}  & \multicolumn{2}{c}{} \\[0.5ex]
				\hline\hline \\[-1ex]
			\end{tabular}
		}
		\smallskip
		\\ \textbf{Notes:} \textbf{[a]} Numbers in parenthesis are one standard deviation and apply to the last significant digits.
		\textbf{[b]} Ref.~\cite{baldacci2005high}. \textbf{[c]} Fixed to the ground state value.
	\end{table*}
	
	\noindent
	Tables~\ref{tab:ground} and~\ref{tab:ex} show a great improvement in the precision of all the 
	spectroscopic parameters of \ce{CH2D^{35}Cl}.
	In particular, due to the first observation of $b$-type transitions, the constants $A$
	and $\Delta_K$ are confidently determined with errors that are three orders of magnitude smaller
	than those reported in Ref.~\cite{baldacci2005high}. Also, all the remaining parameters are one or
	two orders of magnitude more precise.
	
	\noindent
	In the analysis of the $\varv_5=1$ and $\varv_6=1$ states, coupled through a $c$-type Coriolis
	interaction \cite{baldacci2005high}, the vibrational energies $E$ and the resonance parameters
	$G_{c}$, $G_c^{JK}$ and $F_{ab}$ have been determined as well.
	
	\noindent
	As far as the \ce{CH2D^{37}Cl} species is concerned, this work represents the first detailed investigation of its
	rotational spectrum. Generally speaking, the spectral analysis is satisfactory; for instance, the standard deviation
	of the fit ($\sigma=0.77$) indicates that the data set are adequately reproduced within their expected uncertainties.
	Additionally, the obtained spectroscopic parameters have errors similar to those of \ce{CH2D^{35}Cl} and their values
	are consistent with the isotopic substitution. The only exception is represented by $\Phi_{JK}$, whose values in
	\ce{CH2D^{35}Cl} and \ce{CH2D^{37}Cl} have opposite signs. This can be explained by the fact that different sets
	of sextic centrifugal distortion terms have been fitted and the analyses are based on different data-sets.
	
	\noindent
	The complete list of all the observed transitions is deposited as supplementary material.
	
	
	\section{Conclusions}
	
	\noindent
	The rotational spectra of \ce{CH2D^{35}Cl} and \ce{CH2D^{37}Cl}, the singly-deuterated forms of chloromethane,
	have been observed in the millimeter region for the first time.
	Exploiting the Lamb-dip technique, precise rest frequencies have been retrieved for a large range of 
	$J$ and $K_a$ transitions.
	Besides the ground state spectra, about 200 transitions of \ce{CH2D^{35}Cl} in its $\varv_5=1$
	and $\varv_6=1$ excited states have been also recorded.
	The analysis of a conspicuous data-set led to accurate values of many spectroscopic parameters, including
	the rotational constants $A$, $B$, $C$, several centrifugal distortion terms, and quadrupole coupling constants
	$\chi_{ii}$(Cl).
	Overall, the quality of all spectroscopic parameters of \ce{CH2D^{35}Cl} has been improved with the respect to previous works
	\cite{baldacci2005high}. As to \ce{CH2D^{37}Cl}, our set of constants is the first reported in literature.
	
	\noindent
	The main aim of this work is to provide reliable spectral prediction to guide radio-astronomical searches of \ce{CH2DCl},
	a species which might be present in the low-mass protostar IRAS 16293-2422.
	The recent detection of chloromethane in this source has revealed \ce{CH3Cl} to be fairly abundant and to possess a
	rotational temperature ($T_\mrm{rot}$) of \emph{ca.} 100\,K \cite{fayolle2017protostellar}.
	Without astrochemical models, it is hard to guess a reasonable abundance of \ce{CH2DCl}.
	In similar molecules, e.g., methanol and methyl cyanide, deuterium fractionation can vary quite a lot,
	ranging from 90 \% to 4 \% for \ce{CH2DOH} and \ce{CH2DCN}, respectively \cite{parise2002detection,calcutt2018alma}.
	
	\noindent
	However, it is known that deuterium fractionation processes are very efficient at low temperature;
	therefore, one could expect \ce{CH2DCl} to possess a $T_\mrm{rot}$ as low as 10\,K.
	In that case, the spectrum of \ce{CH2DCl} will peak in the 2--3\,mm region, where
	many radio-telescopes (such as IRAM 30m, APEX, and ALMA) offer wide spectral coverage and high-sensitivity.
	In case of a higher $T_\mrm{rot}$, however, \ce{CH2DCl} emission will peak at higher frequencies, e.g.,
	around 400\,GHz at 100\,K.
	ALMA, whose capability has been already demonstrated by numerous detections \cite{mcguire2018first,melosso2019sub},
	represents the best ground-based facility covering such frequency region (thanks to its ALMA Band 8 window)
	that can be used to search for \ce{CH2DCl} signatures in the interstellar medium.
	Even in case of a non-detection, it would be instructive to derive an upper limit for its abundance in order to compare it with those of related species \cite{melosso2018laboratory,melosso2019astronomical} or use it within astrochemical models.

	\noindent
	A second important accomplishment of this work is the determination of an accurate set of ground state
	spectroscopic constants for \ce{CH2D^{37}Cl}, which were not available to date.
	They will provide a good starting point for future analyses of the high-resolution ro-vibrational spectrum
	of \ce{CH2D^{37}Cl}, whose acquisition is in progress in our laboratory.
	
	\section{Acknowledgement}
	
	\noindent
	This study was supported by Bologna University (RFO funds), MIUR (Project PRIN 2015: STARS in the CAOS, Grant Number 2015F59J3R),
	and Ca' Foscari University, Venice (AdiR funds).
	The authors gratefully remember Mr.~A.~Baldan for the preparation of the sample of \ce{CH2DCl}.

	\bibliography{mybibfile}
	
\end{document}